\documentclass[10pt,letterpaper]{article}
\usepackage{amsfonts}
\usepackage{amsmath}
\usepackage{amssymb}
\usepackage{graphicx}
\usepackage{opex3}
\usepackage{cite}

\begin{document}

\title{Goos-H\"{a}nchen-like shift of three-level matter wave incident on
Raman beams}
\author{Zhenglu Duan,$^{1,2\ast}$ Liyun Hu,$^{1}$ XueXiang Xu,$^{1}$ and
Cunjin Liu$^{1}$}
\address{$^{1}$  Key Laboratory of Photo-electronic, Telecommunication of Jiangxi Province, Jiangxi Normal University, Nanchang, 330022, China \\
$^{2}$ Center for Engineered Quantum System, School of Mathematics and Physics, University of Queensland, St Lucia, 4072, Qld, Australia }
\email{$^{\ast}$ duanzhenglu@jxnu.edu.cn}

\begin{abstract}
When a three-level atomic wavepacket is obliquely incident on a "medium
slab" consisting of two far-detuned laser beams, there exists lateral shift
between reflection and incident points at the surface of a "medium slab",
analogous to optical Goos-H\"{a}nchen effect. We evaluate lateral shifts for
reflected and transmitted waves via expansion of reflection and transmission
coefficients, in contrast to the stationary phase method. Results show that
lateral shifts can be either positive or negative dependent on the incident
angle and the atomic internal state. Interestingly, a giant lateral shift of
transmitted wave with high transmission probability is observed, which is
helpful to observe such lateral shifts experimentally. Different from the
two-level atomic wave case, we find that quantum interference between
different atomic states plays crucial role on the transmission intensity and
corresponding lateral shifts.
\end{abstract}




\section{Introduction}

The exit point of a light beam totally reflected by a surface separating two
different refractive index media will experience a lateral displacement
corresponding to the incident point. This phenomenon is called the Goos-H%
\"{a}nchen effect \cite{Goos,Picht} and was first experimentally
demonstrated by Goos and H\"{a}nchen in 1947. However,as early as 270 years
ago, Newton intuitively conjectured it via using light ray reflection from
the surface of a mirror \cite{Newton}. Soon after Goos and H\"{a}nchen's
experiment, Artmann theoretically evaluated the lateral displacement using
stationary phase method starting from Maxwell equations \cite{Artmann}. In
1964, Renard applied energy flux conservation to explain Goos-H\"{a}nchen
effect \cite{Renard}.

Recently its consideration has been extended to cases involving multilayered
structures \cite{multilayer}, left-hand material \cite{lefthand},
absorptive, amplified and nonlinear media \cite{nonlinear}. In fact, the
total reflection conditions are not necessary for the lateral shift as long
as there are appropriate phase changes in the plane wave component,
resulting in the transmitted beam also experiencing such shift \cite%
{transmission}. Furthermore, existence of large and negative Goos-H\"{a}%
nchen shift also has been investigated in some circumstances \cite%
{negative1,negative2,Qamar}. Apart from fundamental research, the Goos-H\"{a}%
nchen effect has application in surface plasmon resonance sensor \cite%
{plasmon}, near-field scanning optical microscopy \cite{Near-field} and
optical waveguide switch \cite{Optical-switch}.

Since the Goos-H\"{a}nchen effect comes from wave interference, one can
expect it to also take place in other physical systems, such as acoustic
waves, plasma and matter waves. Refs \cite{acoustic,acoustic1,acoustic2}
investigate lateral shifts of acoustic wave theoretically and
experimentally; while as early as 1960, Hora derived an expression for Goos-H%
\"{a}nchen shift experienced by a beam of quantum particles \cite{Hora}.
Since then the Goos-H\"{a}nchen effect has been investigated in electron
\cite{electron,electron1} and neutron \cite{neutron,neutron1} cases,
theoretically and experimentally.

Thanks to the rapid development of laser cooling and trapping technology,
one can manipulate the ensemble of ultracold atoms for fundamental and
applied investigation under straightforward laboratory conditions, for
example, atom guiding \cite{Guiding}, reflecting \cite{Reflecting} and
diffracting \cite{Diffracting}, resulting in these advancements in the field
of the atom optics.

The ultracold atomic beam transmited through a laser field \cite{zhang} or a
cavity in one-dimensional (1D) constitution \cite{cavity} has been
investigated. When the atomic beam is obliquely incident on a potential
barrier, Goos-H\"{a}nchen-like effect may emerge. Unlike light waves in
traditional optics, matter waves in atom optics can behave differently, for
the atom has internal electronic structures and matter waves of atoms are
coupled vector waves in the light-induced "medium". We have found that there
are large positive and negative lateral shifts for two-level ultracold
two-dimensional (2D) atomic wavepacket in our previous work \cite{Huang}.
This work investigates what happens when a three-dimensional (3D) $\Lambda $%
-type three-level ultracold atomic wavepacket obliquely impinges on Raman
laser beams.

This paper is organized as follows: In Sec. II, we firstly construct an
atomic optics model describing a three-level atomic wavepacket impinging on
pairs of super-Gaussian Raman laser beams acting as a "medium slab". And
then In Sec. III, we obtain the Goos-H\"{a}nchen-like lateral shifts of
reflected and transmitted waves by expansion of reflection and transmission
coefficients. Sec. IV gives the numerical results for lateral shifts under
blue and red detuned situations and the effect of quantum interference on
the lateral shift of transmitted waves. Finally, in Sec. IV we conclude this
paper.

\begin{figure}[tbp]
\begin{center}
\includegraphics[width=3.3in]{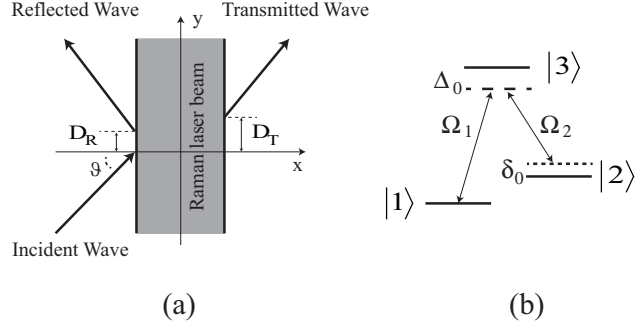}
\end{center}
\caption{{\protect\footnotesize (a) Schematic of lateral shift when a
three-level atomic wavepacket oblique incident on a Raman laser beams. (b)
Energy level structure of atom.}}
\label{fig1}
\end{figure}

\section{Model}

In this work we consider a 3D cold $\Lambda $-type three-level atomic
wavepacket obliquely impinging on a pair of flat top laser fields with a width $L$
(Shown in Fig. 1), which could be realized with a super-Gaussian laser field with a high order \cite%
{supergaussian,supergaussian1,supergaussian2}. The $\Lambda $-type
three-level scheme can be realized, for example, by two $6S_{1/2}$ hyperfine
ground states $F=3$ and $F=4$ (hypefine splitting = 9.19 GHz) of $^{133}$Cs
atoms, which are labeled as $|1\rangle $ and $|2\rangle $, respectively. The
$F^{^{\prime }}=4 $ hyperfine level of the electronically excited state, $%
6P_{3/2}$, forms the intermediate state, $|3\rangle $. The atom interacts
with two laser beams with frequencies $\omega _{L1}$ and $\omega _{L2}$ and
they are detuned from $|1\rangle \rightarrow |3\rangle $ and $|2\rangle
\rightarrow |3\rangle $ transitions by $\Delta $. The Schr\"{o}dinger
equations governing this system are given by

\label{pasi0}
\begin{align}
i\hbar \frac{\partial }{\partial t}\Psi _{1}\left( \mathbf{r},t\right) & =-%
\frac{\hbar ^{2}}{2m}\nabla ^{2}\Psi _{1}\left( \mathbf{r},t\right) -\frac{%
\hbar }{2}\Omega _{1}\left( x\right) e^{-ik_{L1}y}\Psi _{3}\left( \mathbf{r}%
,t\right) \\
i\hbar \frac{\partial }{\partial t}\,\Psi _{2}\left( \mathbf{r},t\right) & =-%
\frac{\hbar ^{2}}{2m}\nabla ^{2}\Psi _{2}\left( \mathbf{r},t\right) -\hbar
\delta _{0}\psi _{2}\left( \mathbf{r},t\right)  \notag \\
& -\frac{\hbar }{2}\Omega _{2}\left( x\right) e^{-ik_{L2}y}\Psi _{3}\left(
\mathbf{r},t\right) \\
i\hbar \frac{\partial }{\partial t}\,\Psi _{3}\left( \mathbf{r},t\right) & =-%
\frac{\hbar ^{2}}{2m}\nabla ^{2}\Psi _{3}\left( \mathbf{r},t\right) -\hbar
\left( \Delta _{0}+\frac{i\gamma }{2}\right) \Psi _{3}\left( \mathbf{r}%
,t\right)  \notag \\
& -\frac{\hbar }{2}\Omega _{1}\left( x\right) e^{ik_{L1}y}\Psi _{1}\left(
\mathbf{r},t\right) -\frac{\hbar }{2}\Omega _{2}\left( x\right)
e^{ik_{L2}y}\Psi _{2}\left( \mathbf{r},t\right)
\end{align}%
In Eqs. (1a)--(1c) we have defined $m$ as the atomic mass, $\nabla
^{2}=\partial ^{2}/\partial x^{2}+\partial ^{2}/\partial y^{2}+\partial
^{2}/\partial z^{2}$, $\Delta _{0}=\omega _{L1}-\omega _{3}$ as the
single-photon detuning and $\delta _{0}=\omega _{L1}-\omega _{L2}-\omega
_{2} $ as the two-photon detuning, $\gamma $ as the decay rate of excited
state $|3\rangle $.

We assume that both laser beams have the same beam profile:

\begin{equation}
\Omega _{1,2}\left( x\right) =\left\{
\begin{array}{lcl}
\Omega _{1,2} & , & 0\leq x\leq L \\
0 & , & \text{\ }x<0,x>L%
\end{array}%
\right. .
\end{equation}

Owing to the Rabi frequencies $\Omega _{1,2}$ being $y$ and $z$ independent,
we eliminate $y$ and $z$ components of the wavefunction through Fourier
transformation

\label{wavefunction}
\begin{align}
\Psi _{1}\left( \mathbf{r},t\right) & =\int dk_{y}dk_{z}\psi _{1}\left(
x,t\right) e^{ik_{y}y-i\hbar \left( k_{y}^{2}+k_{z}^{2}\right) t/2m}, \\
\Psi _{2}\left( \mathbf{r},t\right) & =\int dk_{y}dk_{z}\psi _{2}\left(
x,t\right) e^{i\left( k_{y}+k_{L1}+k_{L2}\right) y-i\hbar \left(
k_{y}^{2}+k_{z}^{2}\right) t/2m}, \\
\Psi _{3}\left( \mathbf{r},t\right) & =\int dk_{y}dk_{z}\psi _{3}\left(
x,t\right) e^{i\left( k_{y}+k_{L1}\right) y-i\hbar \left(
k_{y}^{2}+k_{z}^{2}\right) t/2m}.
\end{align}%
By substituting Eqs. (3a)--(3c) into Eqs. (1a)--(1c) we obtain coupled 1D
Schr\"{o}dinger equations

\begin{equation}
i\hbar \frac{\partial }{\partial t}\,\psi \left( x,t\right) =\left( -\frac{%
\hbar ^{2}}{2m}\frac{\partial ^{2}}{\partial x^{2}}\hat{I}+\hat{V}\right)
\psi \left( x,t\right)  \label{SEQ}
\end{equation}%
where $\psi \left( x,t\right) =\left( \psi _{1}\left( x,t\right) ,\psi
_{2}\left( x,t\right) ,\psi _{3}\left( x,t\right) \right) ^{T}$ is a three
component vector wavefunction, $\hat{I}$ is a $3\times 3$ unit matrix, and $%
\hat{V}$ is the potential as matrix form by

\begin{equation}
\hat{V}=-\frac{\hbar }{2}\left(
\begin{array}{ccc}
0 & 0 & \Omega _{1} \\
0 & 2\delta & \Omega _{2} \\
\Omega _{1} & \Omega _{2} & 2\Delta%
\end{array}%
\right) ,
\end{equation}%
here we have defined effective single-photon detuning and effective
two-photon detuning, respectively, are:%
\begin{align}
\Delta & =\Delta _{0}-\frac{\hbar }{2m}\left( 2k_{L1}k_{y}+k_{L1}^{2}\right)
+\frac{i}{2}\gamma \\
\delta & =\delta _{0}+\frac{\hbar }{2m}\allowbreak k_{y}^{2}-\frac{\hbar }{2m%
}\allowbreak \left( k_{L1}+k_{L2}+k_{y}\right) ^{2}
\end{align}

We consider effective two-photon resonance situation, i.e., $\delta =0$. In
this case, the eigenvalues of the matrix $V$ are $V_{0}=0$ and $V_{\pm
}=-\hbar \left( \Delta \mp \tilde{\Delta}\right) /2$ with $\tilde{\Delta}=%
\sqrt{\Delta ^{2}+\Omega _{1}^{2}+\Omega _{2}^{2}}$, and the corresponding
transformation matrix is

\begin{equation}
U=\left(
\begin{array}{ccc}
\frac{\Omega _{1}}{\Delta -\tilde{\Delta}} & -\frac{\Omega _{2}}{\Omega _{1}}
& \frac{\Omega _{1}}{\Delta +\tilde{\Delta}} \\
\frac{\Omega _{2}}{\Delta -\tilde{\Delta}} & 1 & \frac{\Omega _{2}}{\Delta +%
\tilde{\Delta}} \\
1 & 0 & 1%
\end{array}%
\right)
\end{equation}

\section{Lateral shifts of transmitted and reflected Wavepackets}

The wave packet cannot be described by a stationary wave function, we
therefore address the dynamic evolution of cold atom wavepacket to evaluate
the lateral shift. Here we first take the atom wavepacket in state $%
|1\rangle $ for an example to illustrate how to obtain the lateral shift.
The wave function of incident wavepacket can be described as:

\begin{equation}
\psi _{in}\left( \mathbf{r},t\right) =\int f\left( \mathbf{k}\right) \exp
\left( i\mathbf{k}\cdot \mathbf{r}-iE/\hbar t\right) d\mathbf{k}
\label{incidentwavepacket}
\end{equation}%
where $f\left( \mathbf{k}\right) $ $\left( 2W^{2}/\pi \right) ^{3/4}\exp
\left( -W^{2}\left( \mathbf{k-k}_{0}\right) ^{2}\right) $ is 3D momentum
distribution function around the center wave vector $\mathbf{k}_{0}=\left(
k_{x0},k_{y0},,0\right) $ at initial time. For a 3D wavepacket, the three
components of wave vector $\mathbf{k}$ are decoupled.

After integration, we have

\begin{eqnarray}
\psi _{in}\left( \mathbf{r},t\right) =\left( \frac{1}{2\pi A^{2}}\right)
^{3/4}\exp \left( -\frac{\left( \mathbf{r}-\hbar \mathbf{k}_{0}t/m\right)
^{2}}{4A^{2}}+i\mathbf{k}_{0}\left( \mathbf{r}-\frac{\hbar \mathbf{k}_{0}}{2m%
}t\right) \right)
\end{eqnarray}%
where $A=W\sqrt{1+i\hbar t/\left( 2mW^{2}\right) }$.

The transmitted wave packet can be expressed in a similar way:

\begin{equation}
\psi _{t}\left( \mathbf{r},t\right) =\int Tf\left( \mathbf{k}\right) \exp
\left( i\mathbf{k}\cdot \mathbf{r}-iE/\hbar t\right) d\mathbf{k}
\label{refpacket}
\end{equation}
where $T$ is the transmission coefficient, whose expression is given in the
Appendix.

For a wide momentum distribution $f\left( \mathbf{k}\right) $ of incident
wave, the reflected and transmitted waves will be badly distorted, or even
split, especially around the dip or pole of reflection and transmission
probability, because the "medium slab" severely modifies the reflected and
transmitted momentum distribution far from a gaussian type. In this case,
conventional definition of the lateral displacement in \cite{Artmann} would
be invalid. This situation has been addressed in \cite{negative2,Tamir}. To
avoid severe distortion of the reflected and transmitted waves, a narrow
momentum distribution of incident wave is required. To this end, here we
assume $W$ is sufficiently large that reflection coefficient $T$ is
approximately a constant in the region $k_{0}\pm 1/W$. With the assumption
we can expand the transmission coefficient $T$ around $\left(
k_{x0},k_{y0},0\right) $ up to linear term $T=\left\vert T\left( \mathbf{k}%
_{0}\right) \right\vert \exp \left( \left( \mathbf{\lambda }+i\mathbf{\phi }%
^{^{\prime }}\right) \left( \mathbf{k}-\mathbf{k}_{0}\right) \right) $,
where $\mathbf{\lambda }=\left( d\left( \ln \left\vert T\right\vert \right)
/dk_{x}|_{k_{x0}},d\left( \ln \left\vert T\right\vert \right)
/dk_{y}|_{k_{y0}},0\right) $ and $\mathbf{\phi }=\left( d\phi
/dk_{x}|_{k_{x0}},d\phi /dk_{y}|_{k_{y0}},0\right) $. Hence, the
transmitting wave can be expressed as%
\begin{align}
\psi _{t}\left( \mathbf{r},t\right) & =\left( \frac{1}{2\pi A^{2}}\right)
^{3/4}\left\vert T\left( \mathbf{k}_{0}\right) \right\vert \exp \left( \frac{%
\mathbf{\lambda }^{2}}{4W^{2}}\right)  \notag \\
& \times \exp \left( -\frac{\left( \mathbf{r}^{\prime }-\hbar \mathbf{k}%
_{0}^{\prime }t/m\right) ^{2}}{4A^{2}}+i\mathbf{k}_{0}\left( \mathbf{r}%
^{\prime }-\mathbf{\phi }^{^{\prime }}-\frac{\hbar \mathbf{k}_{0}}{2m}%
t\right) \right)  \label{TW}
\end{align}%
where the wave center wave vector $\mathbf{k}_{0}^{\prime }\mathbf{=k}_{0}+%
\mathbf{\lambda /}\left( 2W^{2}\right) $ and wave packet center $\mathbf{r}%
^{\prime }\mathbf{=r}+\mathbf{\phi }^{^{\prime }}$.

When the center of the transmitted wavepacket traverses through the right
boundary, from (\ref{TW}), it follows that

\begin{align}
y+\phi _{ty}^{^{\prime }}-\hbar k_{y0}^{\prime }t/m& =0 \\
L+\phi _{tx}^{^{\prime }}-\hbar k_{x0}^{\prime }t/m& =0
\end{align}%
And the shift along $y$ direction is

\begin{equation}
D_{t}=\frac{k_{y0}^{\prime }}{k_{x0}^{\prime }}\left( L+\phi _{tx}^{^{\prime
}}\right) -\phi _{ty}^{^{\prime }}  \label{shiftT}
\end{equation}

Considering a large width of the wave packet, $\mathbf{\lambda /}\left(
2W^{2}\right) \approx 0$, Eq. (\ref{shiftT}) becomes

\begin{equation}
D_{t}=\frac{k_{y0}}{k_{x0}}\left( L+\phi _{tx}^{^{\prime }}\right) -\phi
_{ty}^{^{\prime }}  \label{Dt}
\end{equation}%
which is the same as that in \cite{Huang}.

Using the same procedure, we have the lateral shift for the reflected atomic
packet

\begin{equation}
D_{r}=\frac{k_{y0}}{k_{x0}}\phi _{rx}^{^{\prime }}-\phi _{ry}^{^{\prime }}
\label{R}
\end{equation}

Equations (\ref{Dt}) and (\ref{R}) show that we must find the transmission
and reflection coefficients of the atomic wavepacket to determine the
corresponding lateral shifts. The detailed derivation can be found in the
Appendix.

\section{Numerical results and discussion}

In this section we numerically study the lateral shifts using Eqs. (\ref{Dt}%
) and (\ref{R}). In the following we take $k_{x0}=k_{0}\cos \left( \theta
\right) $ and $k_{y0}=k_{0}\sin \left( \theta \right) $, and
correspondingly,
\begin{eqnarray}
\frac{\partial }{\partial k_{x0}} &=&\left( \cos \left( \theta \right) \frac{%
\partial }{\partial k_{0}}-\sin \left( \theta \right) \frac{\partial }{%
k_{0}\partial \theta }\right) \\
\frac{\partial }{\partial k_{y0}} &=&\left( \sin \left( \theta \right) \frac{%
\partial }{\partial k_{0}}+\cos \left( \theta \right) \frac{\partial }{%
k_{0}\partial \theta }\right)
\end{eqnarray}%
where $k_{0}$ is the magnitude of the incident wave vector and $\theta $\ is
the incident angle of atomic wave packet. In such a circumstance the lateral
shift of the reflected and transmitted waves are rewritten as

\begin{equation}
D_{r,t}=-\frac{1}{\cos \left( \theta \right) k_{0}}\frac{\partial \phi _{r,t}%
}{\partial \theta }
\end{equation}

\begin{figure}[tbp]
\begin{center}
\includegraphics[width=3.3in]{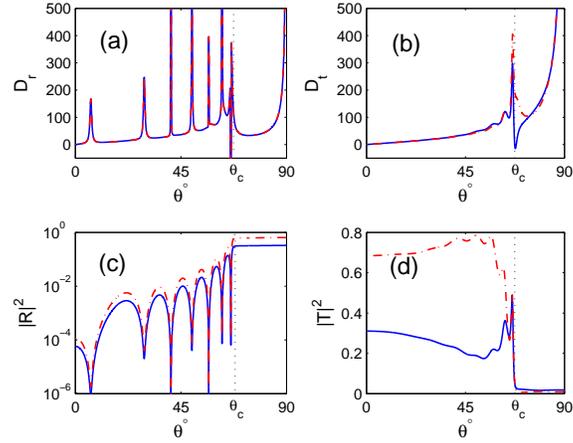}
\end{center}
\caption{{\protect\footnotesize The lateral shift (upper panel) and
reflection(transmission) probability (Lower panel) as a function of incident
angle for atomic waves in state $\left\vert 1\right\rangle $(solid line) and
state $\left\vert 2\right\rangle $(dash-dotted line). The incident atomic
wave is in state $\left\vert 1\right\rangle $ through the blue-detuned laser
beams with different Rabi frequencies. Other parameters are: $L = 30$, $%
\protect\gamma = 0.1$, $\Omega_1 = 2.5$, $\Omega_2 = 3.5$, $k_0 = 0.8$, $k_L
= 0.1$. }}
\label{fig2}
\end{figure}

The variables and parameters in all the figures are dimensionless.

We first consider the blue detuned case with single-photon detuning $\Delta
_{0}=100$. (Here we should stress that, even $\Delta _{0}\gg \Omega
_{1},\Omega _{2}$ and $\gamma $, adiabatic elimination of excited state $%
\left\vert 3\right\rangle $ may lead to loss of many important features of
tunneling and Goos-H\"{a}nchen-like lateral shifts of atomic wave with
internal structure \cite{Duan}.) At this case, the atomic states $\left\vert{%
1}\right\rangle$ and $\left\vert{2}\right\rangle$ can be approximately
expressed as with dressed-state bases (29) and (30):

\begin{eqnarray}
\left\vert 1\right\rangle &\approx &\frac{\Omega _{1}\left\vert
+\right\rangle +\Omega _{2}\left\vert 0\right\rangle }{\sqrt{\Omega
_{1}^{2}+\Omega _{2}^{2}}}  \label{C1} \\
\left\vert 2\right\rangle &\approx &\frac{\Omega _{2}\left\vert
+\right\rangle -\Omega _{1}\left\vert 0\right\rangle }{\sqrt{\Omega
_{1}^{2}+\Omega _{2}^{2}}}  \label{C2} \\
\left\vert 3\right\rangle &\approx &\left\vert -\right\rangle  \label{C3}
\end{eqnarray}%
It can be seen that the scattering properties of atomic waves in states $%
\left\vert 1\right\rangle $ and $\left\vert 2\right\rangle $ are mainly
determined by $\left\vert +\right\rangle $ and $\left\vert 0\right\rangle $
modes. Apparently, if the initial states of the atomic wavepacket is in
states $\left\vert 1\right\rangle $ or $\left\vert 2\right\rangle $, the
atomic wavepacket will mainly "see" a potential barrier since $\mathbf{Re}%
\left( V_{+}\right) >0$. Hence we can define a critical angle as $\theta
_{c}\equiv \cos ^{-1}\left( \sqrt{2m\mathbf{Re}\left( V_{+}\right) /\hbar
^{2}k_{x0}^{2}}\right) $, at which the normal component of kinetic energy of
the incident atomic wave equals the height of the potential barrier. If the
incident angle of the atomic wave is greater than the critical angle, the
atomic wave tunnels through the barrier; otherwise it passes over the
barrier.

Figure 2 shows lateral shifts of reflected and transmitted waves with the
incident atomic wave in state $\left\vert 1\right\rangle $ and $\Omega
_{1}\neq \Omega _{2}$. It can be found that the behavior of reflected and
transmitted waves are quite different in the region $\theta >\theta _{c}$
and $\theta _{c}>\theta >0$. For $\theta _{c}>\theta >0$, the normal
component of kinetic energy of incident atomic wave is greater than the
height of barrier. As a result, the reflected and transmitted waves
oscillate [Figs. 2(c) and 2(d)] and the lateral shifts exhibit giant peaks
at each resonance [Figs. 2(a) and 2(b)]. For $\theta >\theta _{c}$, there
are no peaks on the curves of lateral shifts. In fact, similar phenomena
have been observed in \cite{Huang}. One interesting thing is that, the
lateral shifts of reflected waves in states $\left\vert 1\right\rangle $ and
$\left\vert 2\right\rangle $ are equal, while the transmitted ones are not.
This can be understood that, based on Eqs. (\ref{C1}) and (\ref{C2}), only $%
\left\vert +\right\rangle $ mode contributes to the reflected wave,
consequently the reflected waves in state $\left\vert 1\right\rangle $ and $%
\left\vert 2\right\rangle $ acquire the same phase shift after leaving the
reflection surface and the corresponding lateral shifts are equal although
they have different reflection probabilities [Fig. 2(c)]. While in the
transmission case, states $\left\vert 1\right\rangle $ and $\left\vert
2\right\rangle $ are superpositions of $\left\vert +\right\rangle $ mode and
$\left\vert 0\right\rangle $ mode with different signs, which leads to
constructive or destructive quantum interference. The consequence is that
the transmission intensities of the transmitted waves in state $\left\vert
1\right\rangle $ and $\left\vert 2\right\rangle $ have opposite trends when
the incident angle is varied [Fig. 2(d)]. Also, the lateral shifts are
different [Fig. 2(b)] for different states.

\begin{figure}[tbp]
\begin{center}
\includegraphics[width=3.3in]{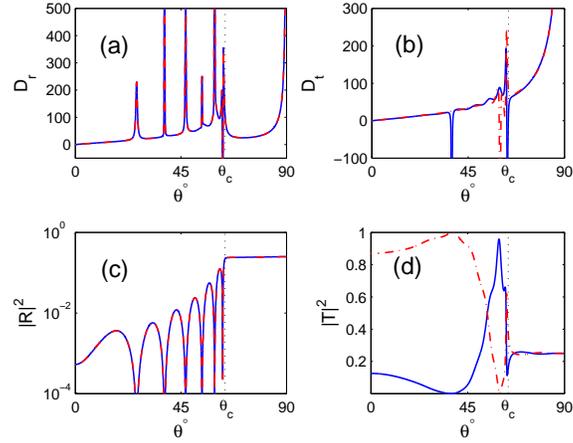}
\end{center}
\caption{{\protect\footnotesize  The lateral shift (upper
panel) and reflection(transmission) probability (Lower panel) as a function
of incident angle for atomic waves in state $\left\vert 1\right\rangle $%
(solid line) and state $\left\vert 2\right\rangle $(dash-dotted line). The
incident atomic wave is in state $\left\vert 1\right\rangle $ through the
blue-detuned laser beams with equal Rabi frequencies. Other parameters are
the same with figure 2. }}
\end{figure}

Now we pay attention to the situation with $\Omega _{1}=\Omega _{2}=3.5$.
Again, the incident atomic wave is set in state $\left\vert 1\right\rangle $%
. Figures 3(a) and 3(b) show the lateral shift of reflected and transmitted
waves in states $\left\vert 1\right\rangle $ and $\left\vert 2\right\rangle $
as a function of the incident angle, respectively. We observe multiple peaks
and dips on the lateral shift curves of the reflected and transmitted waves,
resulting from the resonance scattering. One most interesting things is that
the reflection intensities of atomic waves in states $\left\vert
1\right\rangle $ and $\left\vert 2\right\rangle $ are the same, shown in
[Fig. 3(c)]. Eqs. (\ref{C1}) and (\ref{C2}) are able to explain this
phenomenon, given $\Omega _{1}=\Omega _{2}$, $\left\vert +\right\rangle $
mode equally contributes to the reflected waves in states $\left\vert
1\right\rangle $ and $\left\vert 2\right\rangle $. Another significant
difference is that the lateral shift of transmitted wave in state $%
\left\vert 1\right\rangle $ exhibits a giant negative peak at $\theta
\approx 38^{\circ }$ [Fig. 3(b)]. This corresponds to a nearly vanishing
transmission [Fig. 3(d)], which results from the nearly perfect quantum
destructive interference, instead of resonance scattering.

\begin{figure}[tbp]
\begin{center}
\includegraphics[width=3.3in]{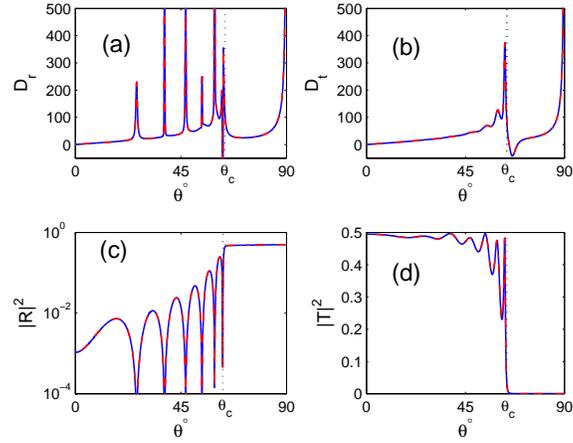}
\end{center}
\caption{{\protect\footnotesize The lateral shift (upper panel) and
reflection(transmission) probability (Lower panel) as a function of incident
angle for atomic waves in state $\left\vert 1\right\rangle $(solid line) and
state $\left\vert 2\right\rangle $(dash-dotted line). The incident atomic
wave is in state $\left( \left\vert 1\right\rangle +\left\vert
2\right\rangle \right) /\protect\sqrt{2}$ through the blue-detuned laser
beams with equal Rabi frequencies. Other parameters are the same with figure
3.}}
\end{figure}

\begin{figure}[tbp]
\begin{center}
\includegraphics[width=3.3in]{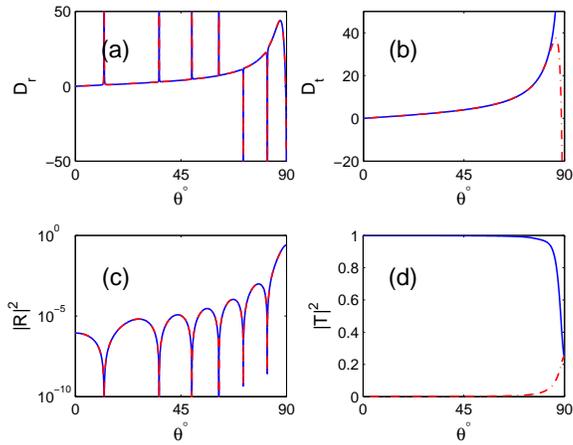}
\end{center}
\caption{{\protect\footnotesize The lateral shift (upper panel) and
reflection(transmission) probability (Lower panel) as a function of incident
angle for atomic waves in state $\left\vert 1\right\rangle $(solid line) and
state $\left\vert 2\right\rangle $(dash-dotted line) through the red-detuned
laser beams with equal Rabi frequencies. the Other parameters are: $L=4$, $%
\protect\gamma =0.1$, $\Omega _{1}=2$, $\Omega _{2}=2 $, $k_{0}=0.8$, $%
k_{L}=0.1$.}}
\end{figure}

Next, we consider another situation when the incident wave is in the
superposition state $\left( \left\vert 1\right\rangle +\left\vert
2\right\rangle \right) /\sqrt{2}$ and $\Omega _{1}=\Omega _{2}$. Now the
atomic wave in the laser beam only contains $\left\vert +\right\rangle $
mode, that is to say, the incident wave only "see" the potential barrier.
Based on the previous analyses, the lateral shifts and reflection
intensities of the reflected and transmitted waves should be the same.
Additionally, the absence of $\left\vert 0\right\rangle $ mode in the
transmission implies no quantum interference happening, therefore the
transmission intensities of the atomic waves in states $\left\vert
1\right\rangle $ and $\left\vert 2\right\rangle $ should also be equal.
Figure 4 verifies these conclusions.

If carefully examining the lateral shifts of transmitted wave in Figs. 2, 3
and 4, one can find that a vanishing transmission intensity is not necessary
to observe a large peak of lateral shift. For example, Fig. 3(b) shows a
large positive lateral shift of atomic wave in state $\left\vert
2\right\rangle $ near the critical angle, while the corresponding
transmission intensity is about $0.6$ [Fig. 3(d)], which is very helpful
to experimentally measure such large Goos-H\"{a}nchen-like lateral shift of
atomic wavepacket.

Finally we study the lateral shifts in the red-detuned case, i.e., $\Delta
_{0}=-25$. Different from the blue detuned case, the scattering properties
of atomic waves in states $\left\vert 1\right\rangle $ and $\left\vert
2\right\rangle $ are mainly determined by $\left\vert -\right\rangle $ and $%
\left\vert 0\right\rangle $ modes, i.e., $\left\vert 1\right\rangle \approx
-1/\sqrt{2}\left( \left\vert -\right\rangle -\left\vert 0\right\rangle
\right) $ and $\left\vert 2\right\rangle \approx -1/\sqrt{2}\left(
\left\vert -\right\rangle +\left\vert 0\right\rangle \right) $. Since Re$%
\left( V_{-}\right) $ is negative for all incident angles, $\left\vert
-\right\rangle $ mode corresponds to a potential well. Lateral shift $%
D_{r,t} $ and intensity $\left\vert R\right\vert ^{2}\left( \left\vert
T\right\vert ^{2}\right) $ of reflected and transmitted waves are displayed
in Fig. 5 as a function of the incident angle $\theta $. One can see that
there are many large positive and negative peaks on the lateral shift curves
of reflected waves in states $\left\vert 1\right\rangle $ and $\left\vert
2\right\rangle $, which are caused by the resonance scattering by the
potential well. The lateral shift of reflected waves in states $\left\vert
1\right\rangle $ and $\left\vert 2\right\rangle $ are equal, which is
similar to the blue detuned case. The reason is that only $\left\vert
-\right\rangle $ mode contributes to the reflected waves. As discussed
before, the quantum interference between $\left\vert -\right\rangle $ and $%
\left\vert 0\right\rangle $ modes leads to unequal lateral shifts and
transmission intensities of transmitted waves in states $\left\vert
1\right\rangle $ and $\left\vert 2\right\rangle $, which is verified by
Figs. 5(b) and 5(d).

\section{Conclusion}

In this work we have investigated the Goos-H\"{a}nchen-like lateral shifts
of a three-level atomic matter wavepacket obliquely impinging on the "medium
slab" made up of two super-Gaussian laser beams. We have obtained the
lateral shifts by expansion of transmission and reflection coefficients.
Results show that the lateral shifts can be either positive or negative
dependent on the incident angle and atomic states. Different from two-level
situation, quantum interference between dressed states manifests itself in
the transmission intensity and the corresponding lateral shift of
transmitted waves, while it has no influence on the properties of reflected
waves.

In particular, we find that there are large lateral shifts with considerable
transmission intensities of atomic wave, which is necessary for high
sensitive measurements of lateral shifts in experiments.

Additionally, the involving the decay of upper level of the atom will result a loss of atom number, hence the total probability flux is less than one.

\section*{Appendix}

Here we present the detailed derivation of the transmission and reflection
coefficients of the three level atom. Considering an atomic plane wave with
incident energy $E=\hbar ^{2}k_{0}^{2}/2m$ and initial value $In=\left(
In_{1},In_{2},In_{3}\right) ^{T}$ incident on the potential. Due to no
coupling between different internal states, the scattering solution for Eqs.
(\ref{SEQ}) outside the laser beam is

\begin{align}
\left(
\begin{array}{c}
\psi _{1} \\
\psi _{2} \\
\psi _{3}%
\end{array}%
\right) =\left\{
\begin{array}{lll}
\left(
\begin{array}{c}
In_{1}e^{ik_{1}x}+R_{1}e^{-ik_{1}x} \\
In_{2}e^{ik_{2}x}+R_{2}e^{-ik_{2}x} \\
In_{3}e^{ik_{3}x}+R_{3}e^{-ik_{3}x}%
\end{array}%
\right) e^{-iE_{x}t/\hbar } & , & x\leqslant 0 \\
\left(
\begin{array}{c}
T_{1}e^{ik_{1}x} \\
T_{2}e^{ik_{2}x} \\
T_{3}e^{ik_{3}x}%
\end{array}%
\right) e^{-iE_{x}t/\hbar } & , & x\geqslant L%
\end{array}%
\right.
\end{align}
where the wave vectors in free space $k_{1}=k_{2}=k_{0}$ and $k_{3}=\sqrt{%
k_{1}^{2}+2m\Delta /\hbar }$. $T=\left( T_{1},T_{2},T_{3}\right) ^{T}$ and $%
R=\left( R_{1},R_{2},R_{3}\right) ^{T}$, respectively, are the transmission
and reflection coefficients of the wave in states $\left\vert 1\right\rangle
$, $\left\vert 2\right\rangle $ and $\left\vert 3\right\rangle $. In the
laser beam, the lights couple the different atomic internal states $%
\left\vert 1\right\rangle $ and $\left\vert 2\right\rangle $ to state $%
\left\vert 3\right\rangle $, described by the coupled equation (\ref{SEQ}).
To diagonalize the coupled equation (\ref{SEQ}), here we introduce three
dressed states

\begin{align}
\left\vert \pm \right\rangle =\frac{\left( \Omega _{1}\left\vert
1\right\rangle +\Omega _{2}\left\vert 2\right\rangle +\left( \Delta \mp
\tilde{\Delta}\right) \left\vert 3\right\rangle \right) }{\sqrt{\Omega
_{1}^{2}+\Omega _{2}^{2}+\left( \Delta \mp \tilde{\Delta}\right) ^{2}}}
\end{align}
\begin{align}
\left\vert 0\right\rangle =\frac{1}{\sqrt{\Omega _{1}^{2}+\Omega _{2}^{2}}}%
\left( \Omega _{2}\left\vert 1\right\rangle -\Omega _{1}\left\vert
2\right\rangle \right)
\end{align}

When we project the coupled equation (\ref{SEQ}) onto this dressed state
basis, the equation decouples. With this preparation, we finally find the
scattering solution in the laser beam taking the form

\begin{align}
\left(
\begin{array}{c}
\psi _{1} \\
\psi _{2} \\
\psi _{3}%
\end{array}%
\right) =U\left(
\begin{array}{c}
A_{+}e^{p_{+}x}+B_{+}e^{-p_{+}x} \\
A_{0}e^{p_{0}x}+B_{0}e^{-p_{0}x} \\
A_{-}e^{p_{-}x}+B_{-}e^{-p_{-}x}%
\end{array}%
\right) e^{-iE_{x}t/\hbar }
\end{align}
where $p_{\pm ,0}=\sqrt{2mV_{\pm ,0}/\hbar ^{2}-k_{0}^{2}}$.

With the continuation condition of wave function $\psi $ and its derivative
at the boundary of the potential, we obtain following equations%
\begin{align}
&\left(
\begin{array}{c}
In_{1} \\
In_{2} \\
In_{3}%
\end{array}%
\right) +\left(
\begin{array}{c}
R_{1} \\
R_{2} \\
R_{3}%
\end{array}%
\right) =U\left(
\begin{array}{c}
A_{+}+B_{+} \\
A_{0}+B_{0} \\
A_{-}+B_{-}%
\end{array}%
\right) \\
&i\left(
\begin{array}{c}
k_{1}In_{1} \\
k_{2}In_{2} \\
k_{3}In_{3}%
\end{array}%
\right) -i\left(
\begin{array}{c}
k_{1}R_{1} \\
k_{2}R_{2} \\
k_{3}R_{3}%
\end{array}%
\right) =U\left(
\begin{array}{c}
p_{+}A_{+}-p_{+}B_{+} \\
p_{0}A_{0}-p_{0}B_{0} \\
p_{-}A_{-}-p_{-}B_{-}%
\end{array}%
\right) \\
&U\left(
\begin{array}{c}
A_{+}e^{p_{+}L}+B_{+}e^{-p_{+}L} \\
A_{0}e^{p_{0}L}+B_{0}e^{-p_{0}L} \\
A_{-}e^{p_{-}L}+B_{-}e^{-p_{-}L}%
\end{array}%
\right) =\left(
\begin{array}{c}
T_{1}e^{ik_{1}L} \\
T_{2}e^{ik_{1}L} \\
T_{3}e^{ik_{3}L}%
\end{array}%
\right) \\
&U\left(
\begin{array}{c}
p_{+}A_{+}e^{p_{+}L}-p_{+}B_{+}e^{-p_{+}L} \\
p_{0}A_{0}e^{p_{0}L}-p_{0}B_{0}e^{-p_{0}L} \\
p_{-}A_{-}e^{p_{-}L}-p_{-}B_{-}e^{-p_{-}L}%
\end{array}%
\right) =\left(
\begin{array}{c}
ik_{1}T_{1}e^{ik_{1}L} \\
ik_{1}T_{2}e^{ik_{1}L} \\
ik_{3}T_{3}e^{ik_{3}L}%
\end{array}%
\right)
\end{align}

For mathematical simplification, here we define the matrixes%
\begin{align}
W& =\left(
\begin{array}{ccc}
e^{p_{+}L} & 0 & 0 \\
0 & e^{p_{0}L} & 0 \\
0 & 0 & e^{p_{-}L}%
\end{array}%
\right) & K& =\left(
\begin{array}{ccc}
ik_{1} & 0 & 0 \\
0 & ik_{2} & 0 \\
0 & 0 & ik_{3}%
\end{array}%
\right)  \notag \\
E& =\left(
\begin{array}{ccc}
e^{ik_{1}L} & 0 & 0 \\
0 & e^{ik_{2}L} & 0 \\
0 & 0 & e^{ik_{3}L}%
\end{array}%
\right) & P& =\left(
\begin{array}{ccc}
p_{+} & 0 & 0 \\
0 & p_{0} & 0 \\
0 & 0 & p_{-}%
\end{array}%
\right)
\end{align}

Then Eqs. (32)--(35) become matrix equations

\begin{align}
&In+R =U\left( A+B\right) \\
&K\left( In-R\right) =UP\left( A-B\right) \\
&U\left( WA+W^{-1}B\right) =T \\
&UP\left( WA-W^{-1}B\right) =KT
\end{align}

By some simple calculation we obtain the transmission coefficient

\begin{align}
T=4F^{-1}KIn
\end{align}

with%
\begin{align}
F& =\left( KU+UP\right) W^{-1}\left( U^{-1}+P^{-1}U^{-1}K\right)  \notag \\
& +\left( KU-UP\right) W\left( U^{-1}-P^{-1}U^{-1}K\right)
\end{align}%
and the reflection coefficient

\begin{align}
R=-G^{-1}DIn
\end{align}
with%
\begin{align}
G& =\left( KU-UP\right) W\left( U^{-1}-P^{-1}U^{-1}K\right)  \notag \\
& +\left( KU+UP\right) W^{-1}\left( U^{-1}+P^{-1}U^{-1}K\right)
\end{align}%
and%
\begin{align}
D& =\left( KU-UP\right) W\left( U^{-1}+P^{-1}U^{-1}K\right)  \notag \\
& +\left( KU+UP\right) W^{-1}\left( U^{-1}-P^{-1}U^{-1}K\right)
\end{align}

\section*{Acknowledgments}

Zhenglu Duan thanks Dr. Zhiyun Hang for helpful discussion and Dr. G. Harris
for his help in manuscript editing. This work is supported by the National
Natural Science Foundation of China under Grants No. 11364021 and No.
61368001, Natural Science Foundation of Jiangxi Province under Grants No.
20122BAB212005.

\end{document}